# DESIGN AND IMPLEMENTATION OF AN IoT BASED LPG AND CO GASES MONITORING SYSTEM

Otoniel Flores-Cortez[1], Ronny Cortez[2] and Bruno González[2]

[1]Department of Applied Sciences, Universidad Tecnologica
de El Salvador, San Salvador, El Salvador
[2]Associate researcher, Universidad Tecnologica de El Salvador,
San Salvador, El Salvador

*ABSTRACT*

*Nowadays use of liquefied petroleum gas (LPG) has increased. LPG is an asphyxiating, volatile and highly flammable gas. In a LPG leak situation, potential health accidents are increased either by inhalation or by combustion of the gas. On the other hand, carbon monoxide (CO) is a toxic gas that comes mainly from combustion in car engines. Breathing CO-polluted air can cause dizziness, fainting, breathing problems, and sometimes death. To prevent health accidents, including explosions, in open or closed environments, remote and real-time monitoring of the concentration levels of CO and LPG gases has become a necessity. The aim of this work is to demonstrate the use of Internet of Things (IoT) techniques to design and build a telemetry system to monitor in real-time the concentration of GLP and CO gases in the surrounding air. To implement this work, as central hardware there is a microcontroller, CO and PLG sensors on the electronic station. Besides, Amazon Web Services (AWS) was used as an IoT platform and data storage in the cloud. The main result was a telematics system to monitor in real time the concentrations of both GLP and CO gases, whose data is accessible from any device with internet access through a website. Field tests have been successful and have shown that the proposed system is an efficient and low-cost option.*

## 1. INTRODUCTION

In Latin American countries liquefied petroleum gas (LPG) is also known as "propane gas", its use has increased in various applications: commercial, industrial and mainly residential [1]. Most home and restaurant kitchens used LPG to cook food on gas stoves. Metal cylindrical tanks were commonly used to store LPG and were placed inside the kitchen or in a nearby location. Another environment where this type of gas is found is within industrial facilities for the production and distribution of the gas itself. LPG is an asphyxiate, volatile and very flammable gas [2]. A leak can occur if it is not stored properly or because of a failure in the storage tank or because of poor handling of the circulation pipes and hoses [3]. In a gas leak situation, possible health accidents inside the house, restaurant or facility increase, either by accidental inhalation or by combustion of the gas [4]. In El Salvador country there is a history of explosive accidents caused by mishandling or leaks of PLG [5] [6] [7] [8]. Carbon Monoxide (CO) is poisonous, taste-, odour- and colourless gas derived from combustion processes of automobiles [9]. CO has fatal consequences if undetected. Intoxication caused by CO is frequent possibly leading to high morbidity and mortality [10]. Symptoms of CO poisoning include dizziness, nausea, weakness, headaches, lethargy, and confusion [11]. Inside garages or closed-door parking lots, residential or commercial, are common places of concentration for CO gas [12]. People that work or user of this kind of spaces are more susceptible to suffer CO intoxication, even if they breathe CO-polluted air by a few minutes [13] [14]. In Latin America health accidents produced by CO gas has increased mainly by lack of monitoring system. [15] [16] [17] [18]. The measurement of LPG and CO gases is obtained in Parts Per Million (PPM) of concentration within the surrounding air. Threshold concentration levels for GLP to take care: 0 to 400 ppm = Normal -- 401 to 800 ppm = Hazardous -- more than 800 ppm = Explosive [19]. For CO gas: 0 to 50 ppm = Normal - 51 to 800 ppm = Dangerous - more than 800 = Deadly [20] [21]. Real-time monitoring for concentration levels of CO and LPG gases within a residential environment or in closed parking lots has become a necessity, as some places where these gases are found grows day by day. In order to help on prevention of health accidents derived from CO and LPG gases, such as poisonings and fires, it is useful to give close or open spaces with a monitoring and early warning system of the concentration levels of these gases. Previous work has related to developing LPG or CO gas monitoring stations. But these have focused on the use of high-cost technological tools [22] [23] [24] [25] or are not connected to a website in real-time [23] [26] [27]. The cloud platforms used for these previous implementation is restrictive or closed-source private or expensive The use of so-called free IoT clouds like Thingspeak or Ubidots is popular among the above studios, but they have some limitations like a limited quantity and frequency of telemetry data. [28] [29] [30] [31] [32]. Some of these previous works only focused only on monitoring one gas either LPG or CO [33] [34] [35]. This work proposed a low-cost electronic system based on IoT technologies, equipped with sensors that are capable of taking a reading of both CO and LPG gases in the surrounding air. The system is also capable of sending sensor data over the Internet and displayed on a website so security staff can aware in real-time possible leaks or high concentrations of these gases and alert users and take measures to avoid possible health damages. Rest of this paper is organized as follows. Section 2 summarizes the prototype development of the IoT system. Section 3 presents experimental results and discussion about the proposal, and Section 4 concludes and present some final comments and ideas to be tackled in future work.

## 2. DEVELOPMENT OF A GAS MONITORING IOT SYSTEM

Methodological development of this proposal system was based on the IoT Architectural Reference Model [36].

### 2.1. Purpose & Requirements Specification

Purpose: automated PLG and CO gases concentration monitoring with Wi-Fi communication and real-time report via a web dashboard. Behaviour: electronics station with sensors capable to take measurement of gases concentration (PPM - part per million) in surrounding air, and a central digital controller programmed to do periodic reading of sensors and send collected data via Wi-Fi to a platform on the Internet. Requirement for management: the system can be monitored via the internet; programming management and configuration of the sensor station can be locally through a USB port provided at the station itself. Requirement for data analysis: the data collected by the sensor is processing in the station itself then send its payload with formatted values to the service in the "cloud". Deployment of applications:  firmware or control software is within microcontroller's flash memory inside the station, to monitor the data produced by the station an IoT platform with a web site dashboard. Security requirements: the system must have basic user authentication for change and access to the IoT platform, however, will be of public access for the visualization of measurements.

## 2.2. Process Specification

Define a single case of operation in a repetitive loop through the firmware in the digital controller: when the system boots, it executes actions to set up internal and external hardware of the microcontroller, then reads the sensors for PLG, CO and Temperature, active an on-board buzzer if reading are above thresholds, finally sends readings to the IoT service through the Wi-Fi network, this entire process is periodic. Figure 1 shows a pseudocode algorithm about described process.

```
Algorithm 1: Algorithm for IoT Station
  Result: periodically send gas concentration data to the IoT platform
  controller hardware initialization;
  WiFi transceiver hardware initialization;
  gas sensors initialization;
  while True do
      read sensors;
      if sensor readings above thresholds then
          active buzzer;
          send sensor data to IoT platform;
      else
          send sensor data to IoT platform;
      end
      wait t;
  end
```

Figure 1. Single case algorithm for electronic station process.

## 2.3. Domain Model Specification

Physical entity: the surrounding air, whose concentration of GLP and CO gases will be read. Virtual entity: represent the physical entity in the digital world. So, we define only one for the surrounding air. Device: programmable digital controller with connected LPG and CO gases sensors. Resource: firmware that runs on the device and a setup script that runs in the IoT cloud. Service - The station service runs natively on the device.

## 2.4. Functional View Specification

The functional view defines functional groups (FG) for the different functions of the IoT system. Each functional group provides functions to interact with instances of concepts defined in the domain model or information related to those concepts. Device FG: includes the programmable controller, PLG and CO gases sensors. Communications FG: protocols used are 802.11 link layer, IPv4 network layer, TCP transport layer, HTTP application layer; to send data payload to the IoT platform system use JSON format. Services FG: There is only one service running within the IoT station controlling service. Administration FG: is performs by the firmware resource. Security FG: security mechanism is a single user credential for IoT cloud configuration. Application FG: the user interface for monitoring the values produced by the IoT system is in the "cloud" as a web page.

## 2.5. Operational View Specification

The Options for deployment and operation of the IoT system are defined. IoT electronic station: mains components are a microcontroller, a Wi-Fi transceiver to internet access, a sensor for LPG gas, a sensor for CO gas. As a visual representation, a led screen for displaying text on the station and a Buzzer to play a sound alarm onsite. Communication API: Amazon Web Services API. Communication protocols: 802.11, IPV4 / 6, TCP and HTTP. Services: controller service hosted on the device written on C programing language and running as a native service. Applications: Web and database Application – AWS web toolbox. Administration: device – Arduino IDE for electronics station and AWS for cloud applications. Figure 2 shows functional blocks for the proposed IoT system.

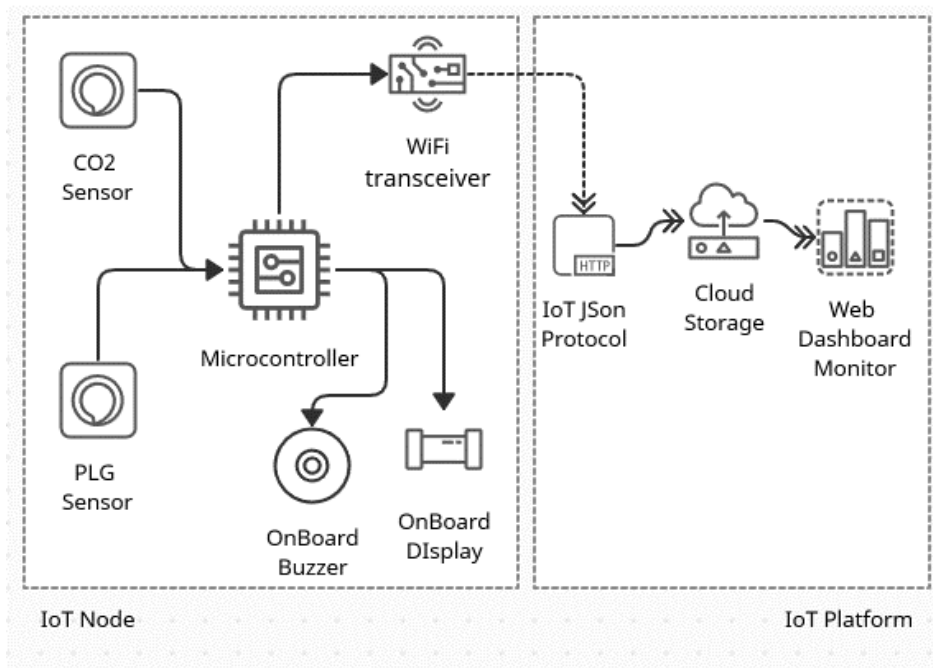

Figure 2: Overview of functional blocks architecture of the proposed IoT system.

## 2.6. Device & Component Integration

Figure 3 shows a schematic diagram with component integration of the IoT electronic station. Mains components used are an Arduino development board for Atmega2560 microcontroller, a ESP8266 chip as Wi-Fi transceiver, a SSD1306 Oled display, MQ-5 LPG sensor, a MQ-9 CO gas sensor and a sound buzzer. Sensors are connected via ADC pins, Wi-Fi transceiver used the UART port pins and I2C port pins are used to handle Oled display, buzzer use one GPIO of microcontroller.

## 2.7. Application Development

The Applications developed to run by the IoT the system are: 1) Device firmware: written in C programming language, the program follows a one loop structure and specific tasks: a. Read sensors values concentrations for PLG and CO gases. b. Store these values locally c. display readings on LCD on device d. Compare readings values to threshold levels of each gas, if current values are above sound buzzer on the station. e. Packet and send the data payload to IoT cloud. g. Wait until the next reading. 2) Service script in the "cloud": developed in JavaScript

Computer Science & Information Technology (CS & IT) 35

language hosted in Amazon Web Services (AWS) cloud. The telemetric protocol JavaScript Object Notation (JSON) is used to send and receive data between sensor station and IoT Platform. AWS services were selected for their low cost, high reliability, and availability versus other similar services. In addition to having a relatively short learning curve. 3) Web application: it was developed using the AWS hosting services, using web-toolbox we setup the web site with data tables and graphical dashboard to display the data generated by the sensor

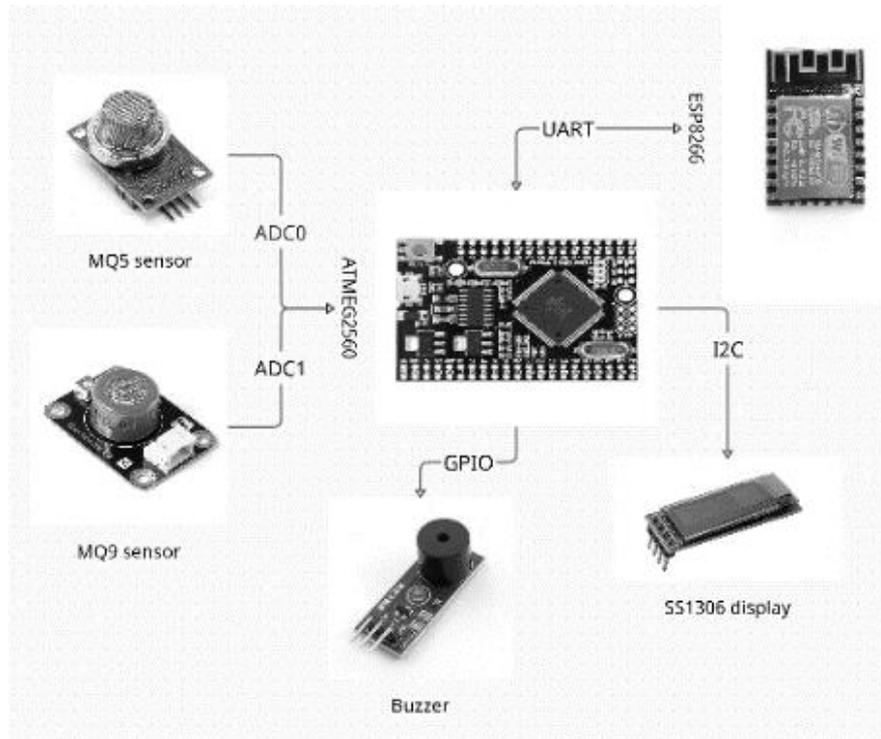

Figure 3: Electronics components integration for sensor station of proposed IoT system.

## 3. RESULTS AND DISCUSSION

### 3.1. IoT monitoring station

One main result of this work was an electronic sensor station prototype that allows take measurements of PLG and CO gases concentrations and send it to the IoT cloud platform, figure 4 shows some photos of the prototype. It is a design that takes on mind needs for a Salvadoran condition, were based on the state-of-the-art, affordable and efficient electronic components. The design of the system allows to add more sensors to the station to increase the gases to be measured. Station reports to the website two values of sensed magnitudes every 10 minutes or when the values raise above threshold levels. The 10 minutes' period is configurable via firmware on microcontroller. Among electrical characteristic of the station prototype, we have: Operation voltage: 110VAC, Current consumption: 0.4W Max, working temperature: +60°C Max. Measurement operation: range PLG and CO 1 to 1000 part per million (ppm) Communication operation: Link: 802.11 Wi-Fi Transmission power: 14 dB tip. Physical installation of station is simple, it can be embedded in a wall of a building or structure, with a height between 1.5 to 2 meters from the ground level. The technical requirements for the place of installation are: access to electric power and Wi-Fi network coverage, the stations are configured for network access by DHCP. The commissioning only requires defining, via the firmware, the network access credentials and the time between reports to the IoT platform, in the case of study it is used 10 minutes.

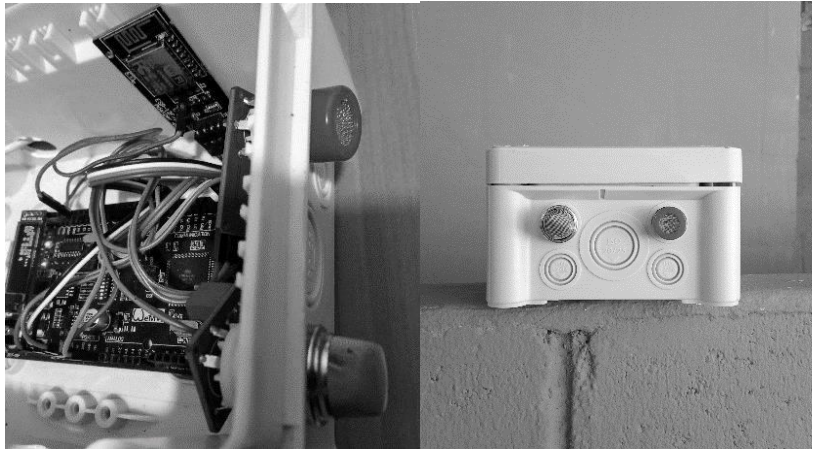

Figure 4: Assembled electronic prototype (left) and outdoors enclosure (right) for the IoT sensor station.

## 3.2. Web Site and Field Test

As a field test designed system was implemented with one node. The station was placed in the center of San Salvador, within the campus of the Universidad Tecnologica de El Salvador, a sector of the capital city with high traffic, specifically between 19 Avenida and Calle Arce. This point was selected to observe its performance and verify operation of the system and to make the necessary adjustments. To monitoring the data collected user can access through any device with Internet access to website with the URL: https://bit.ly/2JBBRkd. This website, figure 5, includes tables and dashboards to view the history of values for PLG and CO gases reported by the station. System performance so far has been satisfactory. The telemetry link has not suffered losses and has remained stable. The website has been available and has not been down. Regarding the data collected, a growth trend has been observed in the concentration of gases that coincides with the rush hours of automotive traffic. It is on early hours of the day and late at night when the tendency on gases concentration is low. Significant effects were observed in gas concentration values in response to climatic conditions such as rainy and windy seasons. The concentration values decrease to their lowest values during the weekdays or after a rain or a gust of wind. This is consistent with the assumption that vehicular traffic is a major source of air pollutant by PLG and CO gases.

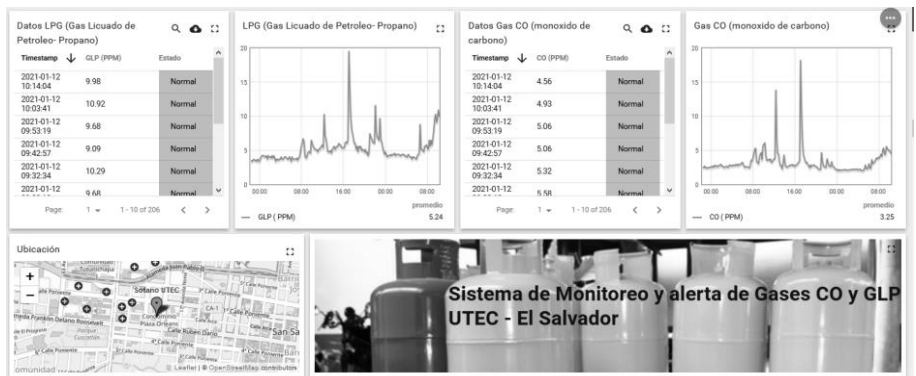

Figure 5: Web site tables and dashboards for collected gases data coming from the sensor station.

## 4. Conclusion and Future Work

Development of an IoT station to monitor the level of LPG and CO gases concentration in surrounding air is a fundamental step for study of behaviour, impacts, and actions for care of the environment, reduction of expenses in health and focus resources on possible solutions for problems that affect quality of human life. The proposed system was developed using state-of-art techniques on electronics, programming, and internet of things, which allowed to produce a piece of low-cost equipment that works according to the expected requirements. Tools such as Atmega microcontroller together with C programming language allows development of efficient IoT prototypes at a low-cost, with short development times and high performance. In addition, using the ready-to-use tools of AWS has allowed fast and simple development of a platform and web site to data monitoring from any device and in real-time. The scientific contribution of this work was to show new and innovative techniques for the use of hardware and software components in the implementation of Internet systems of things. These can be applied in new developments, allowing for fast and efficient prototyping. In the future, this research has the task of developing more stations for different locations within the national territory, conducting validation experiments with additional sensors. In future, we seek to implement a monitoring network through radio frequency links, and analyse massive data or forecasts from the data produced by stations. Also, result of this work can be use in development of new lines of applied research, in areas such as: analysis of aquifers, monitoring in agriculture and livestock fields, analysis of sports performance, etc.

### Acknowledgements

This work was supported by applied research office of Universidad Tecnologica de El Salvador.

## AUTHORS

**Otoniel Flores-Cortez** received his Engineer degree in electrical engineering from the Universidad de El Salvador, San Salvador, El Salvador, in 2005 Since 2005 he works as a lecture and researcher for the electronics department in the Universidad Tecnológica de El Salvador, San Salvador, El Salvador. His research interests include embedded systems, IoT systems.

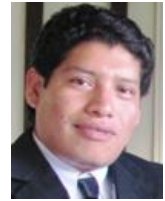

**Ronny Cortez** received his engineering degree in computer science from the Universidad Tecnologica de El Salvador in 2012. Since 2013 is working as a associate professor and research in the computer science department in the same university. His research interests are data mining, cloud computing and machine learning.

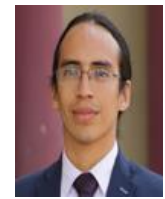

**Bruno Gonzalez** received his engineering degree in computer science from the Universidad de El Salvador in 2012. Since 2011 is working as a professional developer, he worked on different industries like airlines, and services. His research interests are the application of IoT technologies and Machine Learning techniques for natural language processing and computer vision.

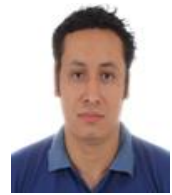